\documentclass[draft]{agujournal2019}
\usepackage{url} 
\usepackage{lineno}
\usepackage[finalnew]{trackchanges} 
\usepackage{soul}
\usepackage{ragged2e}
\usepackage{amsmath}
\draftfalse

\journalname{Space Weather}
\let\oldequation\equation
\let\oldendequation\endequation

\renewenvironment{equation}
  {\linenomathNonumbers\oldequation}
  {\oldendequation\endlinenomath}

\begin{document}

%
%


\title{Opening the Black Box of the Radiation Belt Machine Learning Model}

%
%




\authors{Donglai Ma\affil{1}, Jacob Bortnik\affil{1}, Xiangning Chu\affil{2}, Seth G. Claudepierre\affil{1}, Qianli Ma\affil{1,3}, Adam Kellerman\affil{1},}


\affiliation{1}{Department of Atmospheric and Oceanic Sciences, University of California, Los Angeles, CA, USA}
\affiliation{2}{Laboratory for Atmospheric and Space Physics, University of Colorado Boulder, Boulder, CO, USA}
\affiliation{3}{Center for Space Physics, Boston University, Boston, MA, USA}





\correspondingauthor{Donglai Ma}{dma96@atmos.ucla.edu}




\begin{keypoints}
\item We demonstrate the feature attribution method for a machine learning model of electron flux.
\item We quantify the effects of geomagnetic indices and solar wind parameters on electron flux during a storm time event and a non-storm event. 
\item Our feature importance results identify physical effects that are consistent with our current understanding.
\end{keypoints}

%
%

%
%


\begin{abstract}
\justifying
Many Machine Learning (ML) systems, especially deep neural networks, are fundamentally regarded as black boxes since it is difficult to fully grasp how they function once they have been trained. Here, we tackle the issue of the interpretability of a high-accuracy ML model created to model the flux of Earth's radiation belt electrons. The Outer RadIation belt Electron Neural net (ORIENT) model uses only solar wind conditions and geomagnetic indices as input features. Using the Deep SHAPley additive explanations (DeepSHAP) method, for the first time, we show that the ‘black box’ ORIENT model can be successfully explained. Two significant electron flux enhancement events observed by Van Allen Probes during the storm \change{intervalof}{interval of} 17–18 March 2013 and non–storm interval of 19–20 September 2013 are investigated using the DeepSHAP method. The results show that the feature importance calculated from the purely data-driven ORIENT model identifies physically meaningful behavior consistent with current physical understanding. This work not only demonstrates that the physics of the radiation belt was captured in the training of  our previous model, but that this method can also be applied generally to other similar models to better explain the results and to potentially discover new physical mechanisms. 
\end{abstract}

\section*{Plain Language Summary}
\justifying
A neural network is regarded as a black box model since it can approximate any function but its structure won't give any insights on the nature of the function being approximated. A set of neural network models named OREINT have been developed previously to model the electron flux of the outer radiation belt. In this work, we demonstrate the general flow of explaining the machine learning model of radiation belts and investigate two typical events during the storm and non-storm times. The results identify physically meaningful behavior and are consistent with current physical understanding, additionally providing new insight into radiation belt dynamics. Furthermore, the proposed framework can be generalized for a variety of other machine learning models, including various plasma parameters in the Earth’s magnetosphere.

%
%

%


%
%
%
%

\section{Introduction}

\justifying
The Earth's radiation belts consist of energetic charged particles trapped by the geomagnetic field into two regions, a relatively stable inner zone, and a more dynamic outer zone. These particles range in energy from tens of keV to multiple MeV \cite<e.g.>{van1958observation, vernov1960investigations, lyons1973equilibrium, baker2004extreme, Thorne2010, reeves2016energy, li2019earth}. This radiation environment, exhibiting rich dynamical variations, is known to be particularly hazardous to spacecraft and is difficult to predict, particularly because of the delicate balance between acceleration, transport, and loss, combined with the many different physical processes that produce these effects. To understand outer radiation belt dynamics, the traditional approach involves the integration of the Fokker-Planck (FP) equation \cite<e.g.>{schulz2012particle,ni2008resonant}. Despite being quite successful in capturing the basic structure of the radiation belts \cite<e.g.>{lyons1973equilibrium} or the role played by various waves over long timescales \cite<e.g.>{Thorne2013rapid,qma2016}, the FP model nevertheless suffers from some simplifications which limit its general usage: (1) The diffusion coefficients due to ultra-low frequency waves, whistler-mode waves and electromagnetic ion cyclotron waves are usually parameterized by one of the geomagnetic indices  \cite<e.g.>{ma2015modeling}, but there is no guarantee that such parameterizations are unique and it could not represent nonlinear or any unknown physical process. (2) Artificial boundary conditions are required to drive such simulation when there is no in-situ observation available for a specific event, which could induce simulation errors. To overcome these shortcomings, we present an alternative approach to radiation belt modeling using a set of neural network models to reproduce the radiation belt fluxes. This model, the “Outer RadIation belt Electron Neural neTwork” (ORIENT) model, consists of two sub-models, namely ORIENT-M \cite{2022ma} which covers “Medium” ($\sim50$ keV$-1$ MeV) energies, and ORIENT-R \cite{2021chu} which covers relativistic and ultra-relativistic electron energies ($\sim 1.8-7$ MeV). These models are trained on data from the Van Allen Probes \cite{2013mauk} MagEIS \cite{blake2013magnetic} and REPT \cite{baker2012relativistic} instruments, respectively, and have been extensively tested and validated showing very high accuracy performance for out-of-sample data. Importantly, the ORIENT model successfully captures electron dynamics over long- and short timescales for a range of different energies.

Despite the exciting performance and broad usage of neural network models, the lack of interpretability has been a major concern resulting from the use of such black-box models \cite<e.g.>{camporeale2019challenge}. There has been a growing demand for explainable models in the ML community and as a result, explainable artificial intelligence (XAI) has been developed as a subfield of ML with the goal of providing results with human-interpretable explanations \cite<e.g.>{lipton2018mythos}. Indeed, several interpretable models have been developed recently for forecasting geomagnetic indices \cite<e.g.>{ayala2016modeling, Iong2022}. In this paper, we adapt a state-of-the-art feature attribution method called DeepSHAP \cite{NIPS2017_7062}, to explain the behavior of the ORIENT model at a representative electron energy of $\sim 1$ MeV, during a storm time event and a non-storm time event. The remainder of the paper is organized as follows: In Section 2, we introduce the method of interpreting neural network models used in this study and the general framework to investigate the radiation belt electron flux model. In Section 3, we show the feature attribution results for two significant electron flux enhancement events that occurred during the storm time of 17 March 2013 and the non-storm time of 19 September 2013 and interpret the results in the context of our current physical understanding. These events were originally selected by the Geospace Environment Modeling (GEM) focus group "Quantitative Assessment of Radiation Belt Modeling", and have been well discussed by previous studies \cite<e.g.>{tu2019quantitative,ma2018quantitative}. In Section 4, we conclude with a summary of our key findings.

\section{Methods}

\begin{figure}
\noindent\includegraphics[width=\textwidth]{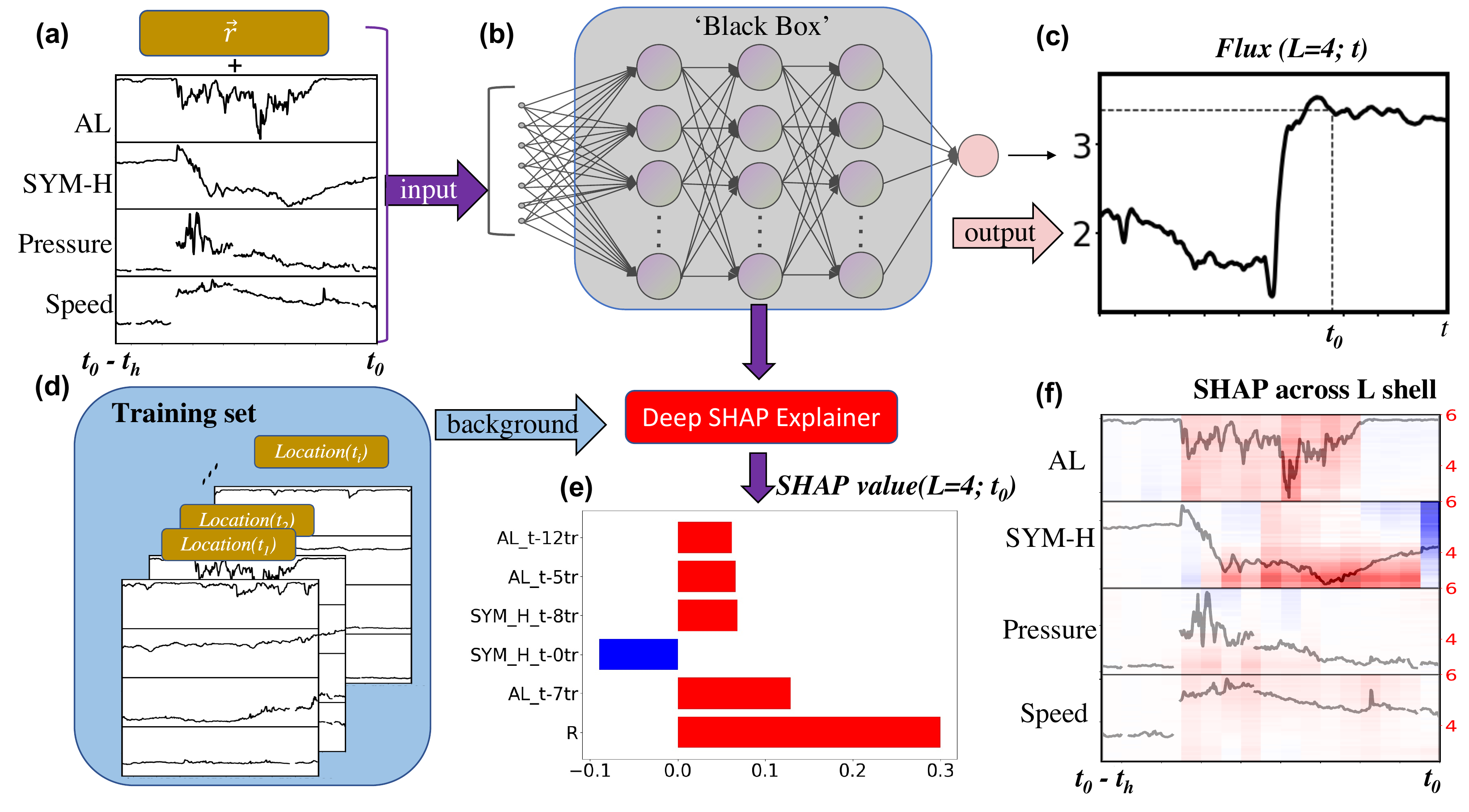}%
\caption{\label{Figure 1} Framework used in explaining the feature importance of the ORIENT model. (a) Model inputs including spacecraft location, geomagnetic indices, and solar wind history, (b) propagation through the neural network model to obtain (c) the output, (e) shows how the feature importance is calculated based on (d) the training dataset and the DeepSHAP Explainer, (f) shows how the feature importance value (SHAP) in (d) is now calculated for the fluxes across all the L-shells and superimposed onto the features, blue corresponds to a feature's decreasing effect on the output and red corresponds to an increasing effect.}
\end{figure}

\justifying
One of the popular and representative methods for machine learning model explanations is SHapley Additive exPlanation (SHAP) proposed by \citeA{NIPS2017_7062}. SHAP assigns each input feature an importance value (SHAP value) for a specific output and the idea of SHAP is based on the Shapley value \cite{shapley:book1952} in game theory. The Shapley value was proposed to determine the {fair} contribution of an individual player in a game with a coalition of players $\mathcal{F}$. {This method can be interpreted as follows: First, consider all the subsets without player $i$ and examine the outcome of each subset $v(\mathcal{S} \mid \mathcal{S} \subseteq \mathcal{F} \backslash\{i\})$. Then, the difference or the marginal contribution is calculated for player $i$: $v(\mathcal{S} \cup\{i\})-v(\mathcal{S})$. Finally, the Shapley value of player $i$ can be computed as:} 
\begin{equation}{}
\begin{aligned}
\phi_i&=\frac{1}{n} \sum_{\mathcal{S} \subseteq \mathcal{F} \backslash\{i\}} \alpha_{\mathcal{S}} \cdot(v(\mathcal{S} \cup\{i\})-v(\mathcal{S}))\\
&=\sum_{S \subseteq F \backslash\{i\}} \frac{|S| !(|F|-|S|-1) !}{|F| !}\cdot(v(\mathcal{S} \cup\{i\})-v(\mathcal{S}))
\end{aligned}
\end{equation}
{where each subset is weighted as number of coalitions excluding $i$ of this subset's size: }$\alpha_{\mathcal{S}}=\left(\begin{array}{c}n-1 \\ |S|\end{array}\right)^{-1}$ {and $n$ is the number of players.}
In brief, the Shapley value is calculated by averaging the expected marginal contribution of one player after all potential combinations have been taken into account. One can transpose the Shapley value into explaining to the machine learning model when the players of the cooperating game become the input features and the profit becomes the output. \citeA{NIPS2017_7062} { show that the Shapely value is a unique solution that can be applied to the entire class of additive feature attribution methods, and they propose SHAP values as a unified measure of feature importance that various methods approximate. The definition of the additive feature attribution method is:}
\begin{equation}
g\left(z^{\prime}\right)=\phi_0+\sum_{i=1}^M \phi_i z_i^{\prime}
\end{equation}
{Here $M$ is the number of simplified features, $z^{\prime} \in\{0,1\}^M$ and  $z^{\prime} $ is a binary variable indication whether feature $i$ is present. Notice that this explanation model use simplified inputs $x'$ that map the original input through the local function $x = h_x(x^{\prime})$ which try to ensure $g\left(z^{\prime}\right) \approx f\left(h_x\left(z^{\prime}\right)\right)$ when $z^{\prime} \approx x^{\prime}$. The unique solution of the explanation model $g$ that satisfies the desirable property (local accuracy, missingness, and consistency) can be found in Equation 8 in} \citeA{NIPS2017_7062}:
\begin{equation}
\phi_i(f, x)=\sum_{z^{\prime} \subseteq x^{\prime}} \frac{\left|z^{\prime}\right| !\left(M-\left|z^{\prime}\right|-1\right) !}{M !}\left[f_x\left(z^{\prime}\right)-f_x\left(z^{\prime} \backslash i\right)\right]
\end{equation}
where the values $\phi_i$ are exactly the Shapley values and $f_x\left(z^{\prime}\right)=f\left(h_x\left(z^{\prime}\right)\right) = f(z_S)$, and $S$ is the set of non-zero indexes in $z^{\prime}$. However, it is difficult to calculate the exact Shapely values because most ML models cannot handle arbitrary patterns of missing input values when calculating $f(z_S)$ ( i.e. for a given neural network model, we can not remove any input features). \remove{SHAP approximates $f(z_S)$ with conditional expectation $E\left[f(z) \mid z_S\right]$ and can be further approximated as $f\left(\left[z_S, E\left[z_{\bar{S}}\right]\right]\right)$ when assuming feature independence and model linearity (Equations 9-12 in ).} \add{To address this issue, $f(z_S)$ is approximated with the conditional expectation $E\left[f(z) \mid z_S\right]$ and the proposed SHAP value is the solution to Equation (3) with such approximation.}

\add{The exact computation of SHAP value is still challenging because of the factorial computational complexity which becomes intractable for models with many features. }\add{For the deep neural network model, they proposed a method to approximate the SHAP values known as Deep SHAP. This method further approximates $E\left[f(z) \mid z_S\right]$ as $f\left(\left[z_S, E\left[z_{\bar{S}}\right]\right]\right)$ when assuming feature independence and model linearity (Equations 10-12 in Lundberg and Lee (2017)). } The interpretation of this assumption is that the missing input feature can be approximated by the average values (expectations) of this feature from the given background samples (i.e. training dataset). Such assumption can be used in Deep SHAP which is a computationally effective method specifically for deep neural network models. Consider a linear model with feature independence as an example: $f(x)=\sum_{j=1}^M w_j x_j+b$, the SHAP results are $\phi_0(f, x)=b$ and $\phi_i(f, x)=w_j\left(x_j-E\left[x_j\right]\right)$. Similarly, when assuming the deep model is linear, one can treat each layer as a linear model and the SHAP values can be linearly backpropagated through the network to get the importance of each feature (see Equations 13-16 in \citeA{NIPS2017_7062}). For each data point $\vec x$ to be explained, the sum of its SHAP values equals the difference between the model prediction $f(\vec{x})$ and the mean value of model prediction of the background samples:
\begin{equation}
    \sum \phi_i =  f({\vec{x}}) - E(f(\vec{x}))
\end{equation}
where $\phi_i$ is the SHAP value of the $i$th feature. \add{Positive (Negative) SHAP value $\phi_i$ indicates the $i$th feature has a positive (negative) impact on the output, leading the model to have a higher (lower) output than the average background $E(f(\vec{x}))$.} A detailed explanation of Deep SHAP can be found in \citeA{chen2021explaining} and \citeA{NIPS2017_7062}.

Here, we open the 'black box' of the ORIENT model based on the idea discussed above and demonstrate the method in Figure 1. Figure 1a-c shows the information flow of the ORIENT-M model which we will interpret. The flow proceeds as follows. To model the flux at any time and any location: $F(\vec{r},t_0)$, we use the time history of geomagnetic indices and solar wind parameters as the features of the model. For the 909 keV channel, the length of the look-back window is 18 days long, obtained after hyperparameter tuning \cite{2022ma}, and the time series are 2-hour averages of AL, SYM-H, $P_{sw}$ and $V_{sw}$ (Figure 1a). Those time series along with the position vector $\vec{r} = (L,\sin(MLT), \cos(MLT), MLAT)$ are then input into the neural network model (Figure 1b) and the output flux $F(\vec{r},t_0)$ is obtained on a log scale (Figure 1c). It is worth noting that in this study we are focusing on the equatorial flux values and because the electron flux at 909 keV has an MLT-symmetric distribution, we only consider L-dependence (setting MLAT and MLT to zero). The detailed description of the ORIENT-M model is given in \citeA{2022ma}. For the flux output at $t_0$ (e.g., $F(L = 4,t_0)$), the SHAP value for each feature is estimated using DeepSHAP and the background data is provided by random samples of 20 percent of the training dataset with $\sim 500,000$ data points. Positive SHAP values (indicating that the feature value contributes to an increase in the output) are colored in red while negative values (indicating that the feature values contribute to a decrease in the output) are colored in blue, as shown in Figure 1e for the single output. We can then calculate the SHAP values at $t_0$ across all L-shells so that the SHAP values across L are represented by $\phi_{Q}(t, L)$ while $t_0 - t_h < t < t_0, 2.6 < L < 6 $ and $Q$ represents one of the input geomagnetic or solar wind parameters. The results are then color-coded (x-axis: time, y-axis: L) as shown in Figure 1f; each $Q$ is also shown as the black lines in the figure so that each SHAP value maps to the corresponding input feature. 

\section{Input Feature Attribution Results}

\subsection{Storm time enhancement}
\begin{figure}
\noindent\includegraphics[width=\textwidth]{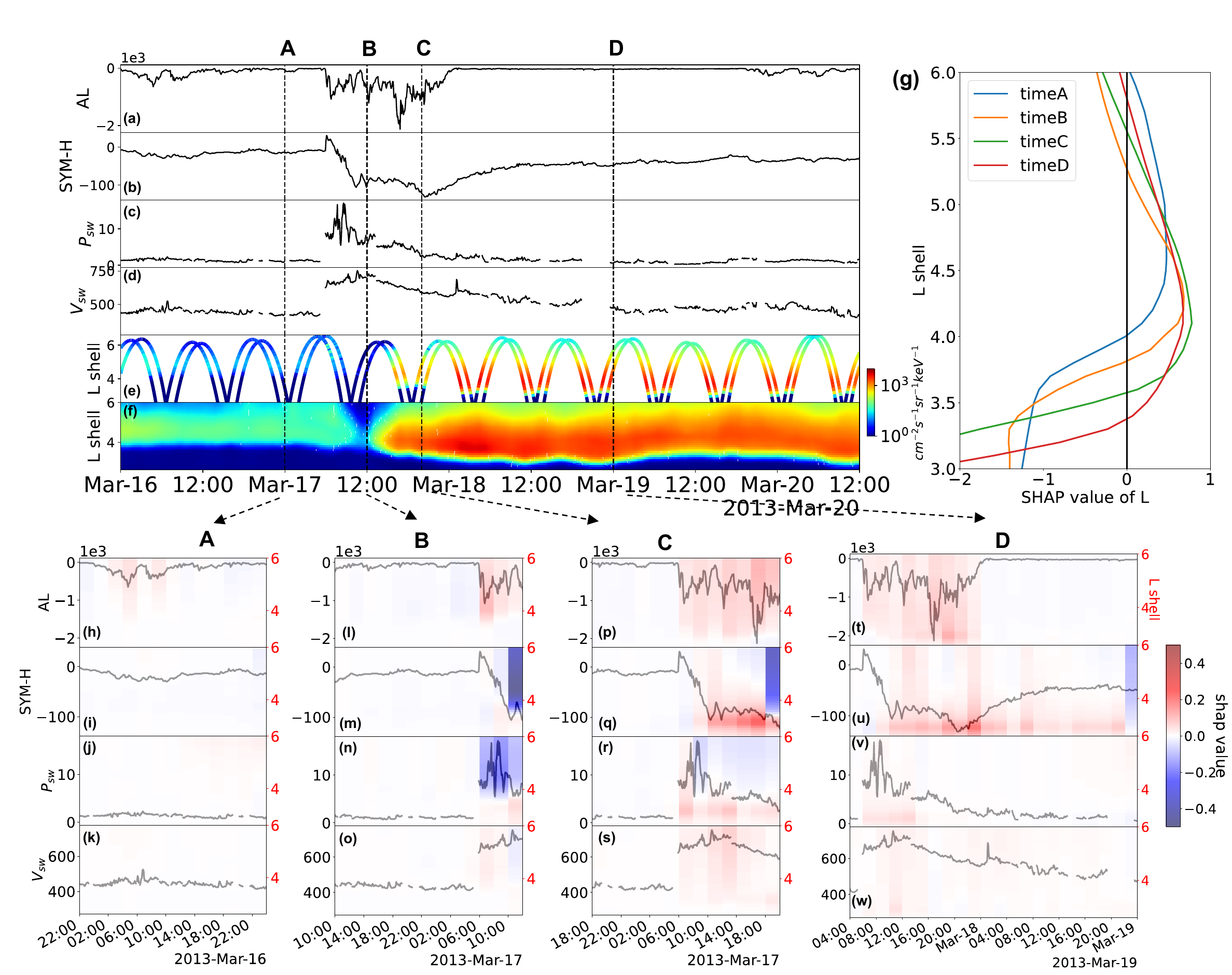}%
\caption{\label{Figure 2} ORIENT Model output and feature attribution results for the storm time event of 17 March 2013. Input time series: (a) AL index, (b) SYM-H index, (c) Solar wind dynamic pressure, $P_{sw}$ (d) Solar wind speed, $V_{sw}$ (e) Observed 909 keV electron fluxes as a function of time and L-shell, (f) ORIENT model reconstruction of 909 keV electron fluxes on the equatorial plane, (g) Feature importance based on L-shell, (h)-(k): Color-coded SHAP feature contributions for the model output at time A (2013-Mar-17-0:00) and the corresponding input \add{zoomed} in a one-day look-back window, (i)-(o): at time B (2013-Mar-17-12:00), (p)-(s): at time C (2013-Mar-17-20:00), (t)-(w): at time D (2013-Mar-19-0:00) but \add{zoomed} in a two-day look-back window.}
\end{figure}

Figure 2 shows the model and SHAP values of input features for the storm-time radiation belt acceleration event that occurred during 17-19 March 2019. The observed flux of 909 keV electrons is shown in Figure 2e as a function of time and L-shell, and the results of the ORIENT-M model on the equatorial plane are shown in Figure 2f which successfully capture the initial dropout and following enhancement during the storm. The SHAP values are then calculated for every feature, at every time in the look-back time series, and for the flux value at every L-shell, and shown in Figure 2h-2w for outputs at four different time snapshots (A-D) as indicated by the vertical dashed lines in Figure 2. Figure 2g shows the SHAP values as a function of L-shell only. It is worth noting that Figure 2h-2w are zoomed to the most recent time history to highlight the storm effect.

The idea of Figure 2h-2w is that each feature, and indeed each sample in the time series of each feature can contribute to the final flux value in different ways. We see that these can change depending on which L-shell the flux value is observed. At time A, before the storm onset, the feature importance of geomagnetic activity and solar wind parameters are all close to zero, indicating that they do not contribute significantly to the flux variation. At time B, a strong dropout event occurs: the SYM-H index gradually decreases to $\sim -100$ nT accompanied by an AL decrease to $\sim - 1000$ nT. The $P_{sw}$ and $V_{sw}$ were notably enhanced at the same time. Despite the fact that these changes took place simultaneously, they have very different contributions to the output: Changes in AL contribute positively while changes in $P_{sw}$ and SYM-H contribute negatively as shown in Figure 2l-2n. Thus, the SHAP values indicate that there is a local competition between acceleration and loss of fluxes at $L>4$. The contributions from the $V_{sw}$ are very close to zero compared with the other three parameters as shown in Figure 2o. Interestingly, the feature contributions for the dropout period are only significant at $L > 4$ for time snapshot B.

At time snapshot C, the feature contributions to this enhancement are very different from previous times: The contributions from AL are positive and extend through the whole L range as shown in Figures 2p and 2t. The contributions from SYM-H are positive at the lower L-shells but the importance closest to time C is negative for the higher L-shells as shown in Figure 2q indicating that it contributes to short-term loss at higher L, but long-term acceleration at the lower L-shells. The contributions from $P_{sw}$ are less important compared to time snapshot B. Remarkably, the change of pressure at the same time gives negative contributions at higher L-shell and positive ones at lower L-shells as shown in Figure 2r, suggesting that the same parameter can drive different responses at different L-shells. The feature importance of the solar wind is positive and larger than at time B. 

At time D, the contributions from pressure and solar wind are again close to zero as shown in Figures 2v and 2w.
Figure 2g demonstrates that the SHAP values as a function of L are always positive at the acceleration region ($L\sim 4-4.5$) and negative at lower L-shell, suggesting an ambient, low-level acceleration which is presumably just outside the plasmapause, and an ambient loss outside of $L\sim 6$, presumably due to an outward drift to the magnetopause boundary. Interestingly, the contribution at $L \sim 6$ is around zero at time A but turns negative at time B when the dropout happens, possibly due to enhanced radial outward drift \cite{shpritsoutward}. It is also worth noting that the intersection of zero SHAP value and the contributions ($L \sim 4$) moves to the lower L-shells as the storm progresses, suggesting that the plasmapause and acceleration region (due to chorus waves) moved to lower L-values \cite<e.g.>{Thorne2013rapid}.

The SHAP technique is an additive feature attribution method, meaning that the contributions for each value of $Q$ could be added over the look-back window to compare their total contributions to the flux during this storm time event. This is illustrated in Figure 4 for two events and two L-shells in each event. Figure 4a shows the March 2017 storm described above, giving the modeled flux at a relatively high L-shell ($L = 5.8$) and 4b shows the 
sum of SHAP values for each $Q$ together with the SHAP value of L-shell as a function of time. Figure 4c and 4d show the same results but for $L = 3.6$. For high L-shell, as the storm progresses, the SYM-H and $P_{sw}$ firstly give negative contributions to the output, and then recover to zero while the $V_{sw}$ and AL have positive contributions. At the dropout time around March 17-12:00, the negative contribution from SYM-H, $P_{sw}$, and $L$ dominate. For low L-shell, the SYM-H and $P_{sw}$ both try to enhance the flux at first, but the positive feature contribution from pressure quickly recovers to zero while the contribution from SYM-H is found to be predominantly rising. The contribution from AL also rises but is less than SYM-H at the lower L-shell. Interestingly, at the end of the storm (March 20), the contribution to low L-shell enhancement is dominated only by the SYM-H index. The fact that the intersection in Figure 2g moves downward to the lower L-shell during the storm also plays an important role in the flux enhancement, which is indicated by the enhancement of SHAP value from the L-shell as shown in Figure 4d.

The feature attribution results are in consistent with our physical understanding and previous studies during this storm time event \cite<e.g.>{turner2013storm,ma2018quantitative}. The dayside magnetopause causes trapped electrons to escape to the system's outer boundary through magnetopause shadowing \cite{ukhorskiy2006storm} and occurs as a result of the magnetopause moving suddenly inward in reaction to increased solar wind dynamic pressure. The subsequent enhanced outward radial transport by enhanced ultra-low frequency wave activity can facilitate the sudden dropouts of electrons \cite{shpritsoutward}. Similarly, our feature attribution results show that the increased pressure during the storm intends to drastically reduce the flux, but at only a high L-shell. Furthermore, simultaneous changes in AL and SYM-H lead to very different outcomes. The cluster of AL peaks always contributes to the acceleration of fluxes at higher L-shell while the main phase of SYM-H contributes to the dropout at higher L-shell and enhancement at lower L-shell, possibly due to the so-called 'Dst effect' \cite{dsteffect}. The ring current alters the magnetic field and electrons moved radially outward (inward) to conserve their third adiabatic invariant, their fluxes decrease (increase) for fixed energy as their first invariant is also conserved. One of the difficulties in analyzing geoeffectiveness during storms is that all the responses happen simultaneously and the driver is a combination of different effects. Thus, with the feature attribution method introduced here, we can quantitatively analyze and unravel the contributions from different inputs to the system.


\subsection{Non-storm time enhancement}
Figure 3 shows a similar analysis to Figure 2 but examines a non-storm time enhancement event that occurred during the period 18-21 September 2013, despite the fact that SYM-H remained higher than $-30$ nT (which would typically signal a geomagnetically '`quiet' period). The electron fluxes gradually grew near the  $5 < L < 6$ region and the ORIENT model successfully reproduced this behavior on the equatorial plane as shown in Figure 3e-3f. The enhancement at times B, C, and D exhibit a clear correlation to the peaks in AL with positive SHAP values, and remarkably, the strongest peak where AL$ \sim -1000 $ nT was found to coincide with the highest feature contributions values (deep reds). In contrast, the period of AL $\sim 0$ nT gives generally weak negative contributions to the fluxes as shown in Figures 3l, 3p, and 3t. {Interestingly, the positive SYM-H in the last 2-hour window of time A gives positive contributions to the flux, and at times B and C, they give negative contributions. The reason for these contradictory results is not clear and needs further investigation.} Figure 3g demonstrates that the contribution from higher L-shells becomes gradually enhanced from time A to time D. Interestingly, the intersection of zero SHAP value and contributions around $L \sim 4.3$ does not move which is a very different response than that observed in the storm time event (Figure 2). The summed SHAP values of each $Q$ and the SHAP of L are shown in Figure 4f ( for $L = 5.2$) and 4h ( for $L = 3.2$). For the high L-shell case, it is seen clearly that there was a sudden enhancement in the AL contribution which results in a major increase in flux as shown in Figure 4f. For the low L-shell case, the contributions are much less significant than at the high L-shell since the fluxes are extremely low throughout the time interval displayed. 

The acceleration during the non-storm time enhancement has been discussed in previous studies \cite{schiller2014nonstorm,su2014nonstorm}. By performing 3D radiation belt simulations, \citeA{ma2018quantitative} showed that radial diffusion by ultra low-frequency waves plays a dominant role in this enhancement. However, none of these studies made a direct connection between the acceleration event and substorm activity. Here, we show clear evidence that this non-storm time enhancement is related to the substorm activity. Along with the result of the storm time event, the feature attribution method shows that the substorm injection process may contribute to the acceleration of electron flux, which is consistent with the acceleration scenario presented in \citeA{jaynes2015source}.

\begin{figure}
\noindent\includegraphics[width=\textwidth]{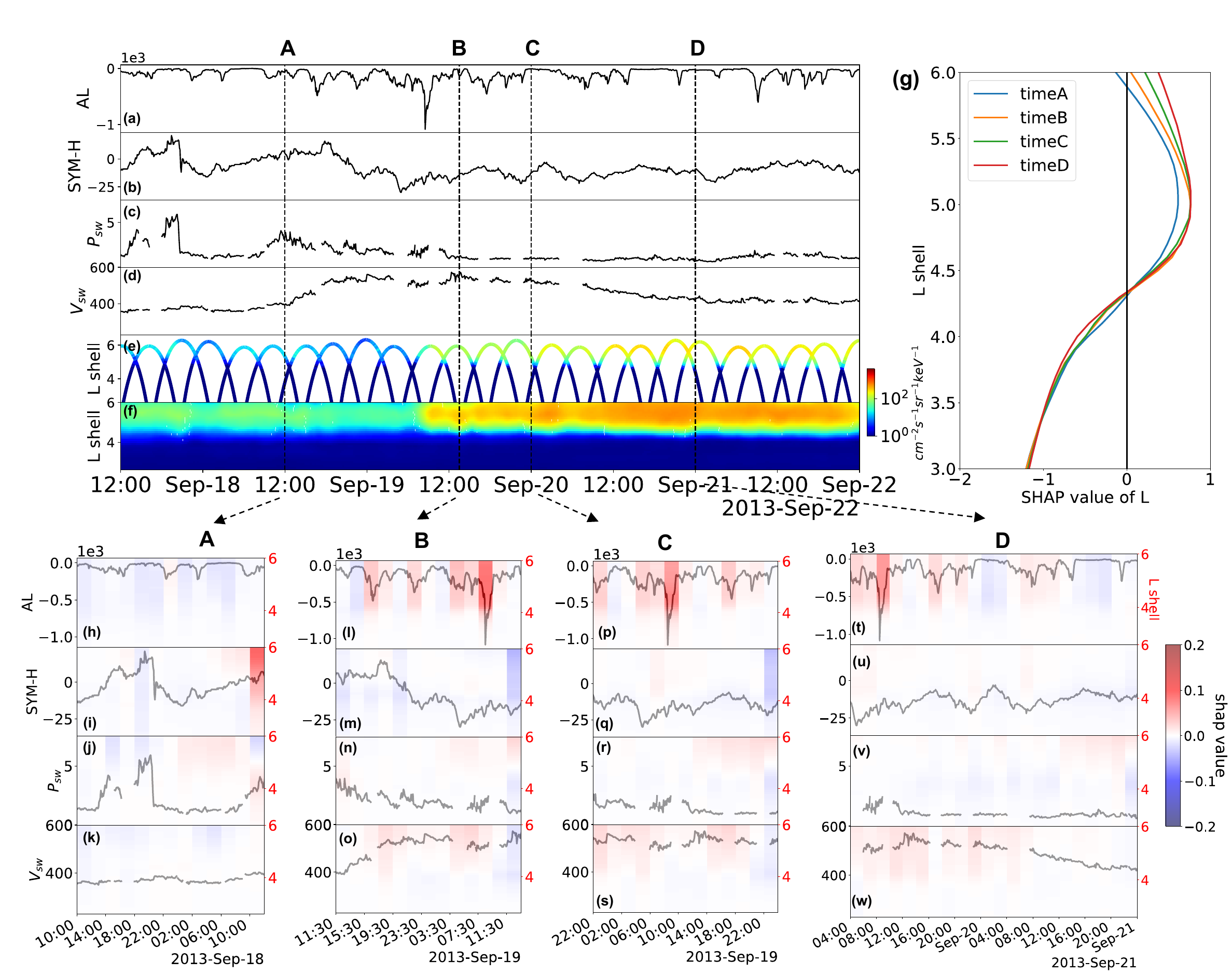}%
\caption{\label{Figure 3} Similar analysis to Figure 2 but for a contrasting non-storm time radiation belt acceleration event, the selected times A: 2013-Sep-18-12:00, B: 2013-Sep-19-13:30, C: 2013-Sep-20-0:00 and D: 2013-Sep-21-0:00. Results show the direct influence of injection activity (as proxied by the AL index) to the flux enhancement.}%
\end{figure}

\begin{figure}
\noindent\includegraphics[width=\textwidth]{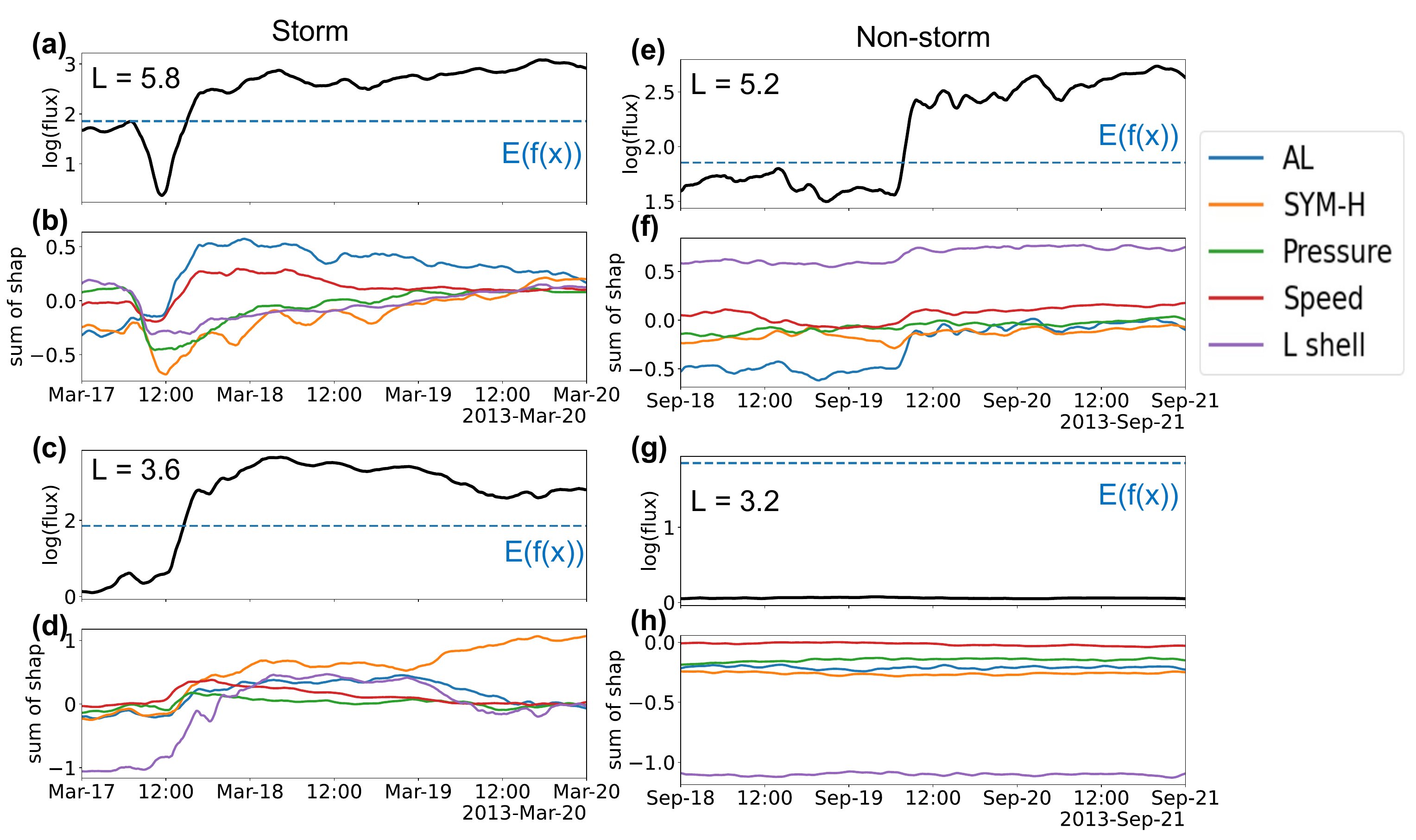}%
\caption{\label{Figure 4} Model results at specific L-shell and the sum of SHAP value for each parameter. (a) Storm time model result at $L = 5.8$, blue dashed line is the baseline value $E(f) = 1.854$ which is the average output of selected background samples. (b) At $L = 5.8$, the sum of SHAP values of AL, SYM-H, Pressure, and Speed, and the SHAP value of L-shell as a function of time. (c)-(d) Same with (a)-(b) at $L = 3.6$. (e-f) Same with (a)-(d) at $L = 5.2$ and $L = 3.2$ for non-storm time. }
\end{figure}

\section{Conclusions and Discussions}
The traditional approach to understanding energetic electron fluxes in Earth’s radiation belt is to use Fokker-Planck simulations with assumed boundary conditions (typically guided by observations) and different, highly parameterized diffusion coefficients. This approach has several limitations, as outlined in the introduction section. In this study, we use a recently developed neural network `ORIENT' model with inputs that include the time history of geomagnetic indices and solar wind parameters. Although the ORIENT model results show high accuracy and can capture the electron dynamics across the energy range (10s keV to several MeV) over long- and short-time scales, the neural network is nevertheless a 'black box' model. The interpretability of such models is an important issue that needs to be addressed and is of interest to all physicists. Here, we apply the feature attribution method based on the DeepSHAP technique to our ORIENT-M model at representative energy (909 keV) and demonstrate the general flow of opening the `black box' of radiation belt (and similar) ML models.

With the proposed framework, we analyze two GEM-challenge enhancement events on 18 March 2013 (storm time) and 19 September 2013 (non-storm time). The DeepSHAP method successfully and quantitatively revealed the feature attribution for the different inputs. For the storm time event, the strong enhancement of solar wind pressure contributed to the rapid dropout seen at higher L-shells, which is consistent with the magnetopause shadowing effect and outward radial diffusion process. The acceleration of electron flux at higher L-shell was contributed dominantly by clusters of AL peaks while at lower L-shell, the acceleration was mainly contributed by the SYM-H index. The rapid decrease of the SYM-H index also contributed to the dropout at high L-shell during storm time. Different contributions to the fluxes from SYM-H at high and low L-shell were seen to be consistent with the well-known 'Dst effect'. Regarding the non-storm time event, the acceleration was found to be clearly correlated to the substorm injection process. These findings, which are consistent with current physical understanding, not only demonstrate the reliability of the interpretation method, but its \change{ability to point to the discovery of potentially missing physical processes}{potential to help the discovery of missing physical processes}. Additionally, the analysis presented demonstrates not only the accuracy of our ML model but most importantly that most of the physical processes are captured in the training of the ORIENT model. Our study thus provides the framework and encouragement for a new way to model and explain radiation belt dynamics and other similar ML models. {For a trustable machine learning model which predicts a physics quantity, we need to make predictions in a physically consistent manner. However, \remove{due to the limitations of satellite data, it is difficult to construct a rich dataset that can cover various extreme events. Moreover, }it is hard to tell just from the matrix, such as mean-square error or correlation coefficient\add{ calculated from years of data}, that the model is good enough. In this paper, we demonstrate that with the proposed SHAP framework, the notable physical process revealed by our previous machine learning model is consistent with our physical understanding. This can be strong evidence that our machine learning model product is deliverable.}

There are many other explanation methods available to compute feature importance for any black box model such as the Local Interpretable Model-Agnostic Explanations (LIME) \cite{ribeiro2016should} and the Integrated Gradient \cite{sundararajan2017axiomatic} methods, among others. \remove{To further check the reliability of our method, we use the Expected Gradient method  which is an extension of Integrated Gradient to compute the feature attribution of the same events, and the results are seen to be very similar to the results presented above (not shown here).} Many of these types of methods define some baseline of missingness which is the key to interpretability because we always want to know if a feature is missing, what would be the effect on the output. We did not discuss the uncertainty of SHAP since we use a large number of background samples $( \sim 100,000)$ to model the missingness and the value would converge to the SHAP values. {The uncertainty mostly comes from the model itself, one can bootstrap the training dataset with sampling and train the model on every bootstrap and then estimate the SHAP value to achieve the confidence level. And this topic is reserved for future work.} However, caution should be exercised when using feature attribution techniques like SHAP. The DeepSHAP and other methods usually assume feature independence so this property not only requires us to develop highly accurate models but also to choose input features with care. Fortunately, we chose input features based on the strategy of adding the most informative predictors sequentially when building ORIENT model. This strategy helps in avoiding features with high co-linearity such as AL and AE. Another important problem is that since we assume the independence of each feature, the hidden interaction between geomagnetic indices and solar wind parameters is ignored. One potential solution is using tree-like models when building the machine learning model, and the SHAP values are able to handle the tree-like model with the global interpretation and feature interaction, which is a topic to be investigated in future studies.

%






\section*{Acknowledgments}
The authors would like to thank the NASA SWO2R award 80NSSC19K0239 for their generous support for this project (as well as subgrant 1559841 to the University of California, Los Angeles, from the University of Colorado Boulder under NASA Prime Grant agreement 80NSSC20K158). XC would like to thank grant NASA ECIP 80NSSC19K0911. XC and QM would like to acknowledge the NASA grant LWS 80NSSC20K0196. JB acknowledges support from the Defense Advanced Research Projects Agency under Department of the Interior award D19AC00009. 

\section*{Open Research}

The data and model files are available at \url{https://doi.org/10.5281/zenodo.6299967} and example code is available at \url{https://github.com/donglai96/ORIENT-M}.


%
%



\bibliography{agusample.bib}

%
%
%
%
%

\end{document}


%
%


\title{Supporting Information for "Opening the Black Box of the Radiation Belt Machine Learning Model"}
%
%

%
%



\authors{Donglai Ma\affil{1}, Jacob Bortnik\affil{1}, Xiangning Chu\affil{2}, Seth G. Claudepierre\affil{1}, Qianli Ma\affil{1,3}, Adam Kellerman\affil{1}}


\affiliation{1}{Department of Atmospheric and Oceanic Sciences, University of California, Los Angeles}
\affiliation{2}{Laboratory for Atmospheric and Space Physics, University of Colorado Boulder, Boulder, CO, USA}
\affiliation{3}{Center for Space Physics, Boston University, Boston, MA, USA}

%
%

%

\begin{article}

%
%

\noindent\textbf{Contents of this file}
\begin{enumerate}
\item Text S1

\end{enumerate}

\noindent\textbf{Introduction}


\noindent\textbf{Text S1.}
%


\noindent\textbf{Data Set S1.} 


\noindent\textbf{Movie S1.} 


\noindent\textbf{Audio S1.} 
\end{article}
\clearpage